\documentclass{epl}
\newcounter{saveeqn}

\newcommand{\alpheqnc}{\setcounter{saveeqn}{\value{equation}}%
     \stepcounter{saveeqn}\setcounter{equation}{0}%
     \renewcommand{\theequation}%
      {\mbox{\arabic{saveeqn}\alph{equation}}}}
\newcommand{\reseteqnc}{\setcounter{equation}{\value{saveeqn}}
     \renewcommand{\theequation}{\mbox{\arabic{equation}}}}

\def\e{{\rm e}}   % Euler's number
\def\d{{\rm d}}   % Differentials
\usepackage{amssymb}
\title{Elasticity of highly cross-linked random networks}
\author{Stephan Ulrich\inst{1} \and Xiaoming Mao\inst{2} \and Paul M. Goldbart\inst{2} \and
  Annette Zippelius\inst{1}}
\shortauthor{Ulrich \etal}
\shorttitle{Elasticity of highly cross-linked random networks}
\institute{                    
  \inst{1} Institut f\"ur Theoretische Physik, Georg-August-Univerist\"at G\"ottingen - \\
  37073 G\"ottingen, Germany\\
  \inst{2} Department of Physics, University of Illinois at Urbana-Champaign - \\
  1110 West Green Street, Urbana, Illinois 61801-3080, USA
}

\begin{document}

\maketitle

\begin{abstract}
Starting from a microscopic model of randomly cross-linked particles with quenched disorder, we calculate the Laudau-Wilson free energy ${\cal S}$ for arbitrary cross-link densities. Considering pure shear deformations, ${\cal S}$ takes the form of the elastic energy of an isotropic amorphous solid state, from which the shear modulus can be identified. It is found to be an universal quantity, not depending on any microscopic length-scales of the model.
\end{abstract}

\section{Introduction}
It is well known that randomly cross-linked networks of molecules
undergo a transition from a fluid to an amorphous solid state, as the
number of cross-links is increased. This sol-gel transition has been
studied in detail \cite{DeamEdwards,Ball,RandomXlinkReview,Panyukov} in recent years, and the structure as well as the
elasticity\cite{GoldstoneFluctEPL,GoldstoneFluct} is well
understood. However, almost all theoretical studies have focussed on the \emph{near-critical} region, 
where the analysis simplifies due to the existence
of a small parameter - the distance to the critical point or, equivalently, the order
parameter.  In contrast, the \emph{highly cross-linked} regime has hardly been
investigated\cite{FiniteDGelation}, yet it is particularly interesting in the application of
random network models to glasses\cite{Zallen}.
   
In this paper we consider a particularly simple network which is built
from spherical particles, connected by harmonic springs. This allows
us to access the highly cross-linked regime and compute the shear
modulus for arbitrary cross-link concentration. We find that the shear
modulus is independent of all microscopic length and energy scales
of the model, such as the excluded-volume interaction, the spring
constant and the length scale that characterizes the localisation of
particles in the amorphous solid state. Instead, the shear modulus is
completely determined by the density of cross-links or, equivalently, the gel-fraction.

\section{Model: randomly cross-linked particles}
We consider a system of $N$ identical particles at positions $\{ {\bf
  R}_1,..., {\bf R}_N \}$ in a $d$-dimensional volume $V$. Permanent
cross-links connect $M$ randomly chosen particles, such that a particular cross-link realization ${\cal C}$ is specified by a list of $M$ pairs of monomers ${\cal C} = \{(i_1,j_1), ..., (i_M,j_M)\}$. The cross-links are modelled by \emph{harmonic
  springs}:
\begin{equation}
  H_{\rm Xlink} = \frac{1}{2a^2} \sum_{e=1}^M ({\bf R}_{i_e} - {\bf R}_{j_e})^2 \;.
\label{HXlink}
\end{equation}
Here, the parameter $a$ can be interpreted as the typical length of a
cross-link. (Energies are measured in units of $k_{\rm B} T$.)
Furthermore, a repulsive excluded-volume interaction $H_{\rm ev}$
is introduced to prevent the system from collapsing. The overall
Hamiltonian is $H = H_{\rm Xlink} + H_{\rm ev}$.
All thermodynamic properties, including the elastic ones, can be
obtained from the partition function 
$ Z({\cal C})=\int \d^d \! R_1 \cdots \d^d \! R_N \, e^{-H}$
which depends on the quenched disorder ${\cal C}$ that specifies the
configuration of cross-links. 

The average over all cross-link configurations ${\cal C}$ is called the \emph{disorder average} and denoted \hbox{by $[ \, \cdot \, ]$}. A realistic choice of the cross-link distribution is the Deam-Edwards distribution \cite{DeamEdwards}:
\begin{equation}
\label{PDeamEdwards}
P_{\rm DE}({\cal C}) \propto \frac{1}{M!}\left( \frac{V}{2 N}
 \frac{\mu^2}{a^d (2\pi)^{d/2} } \right)^M Z({\cal C}).
\end{equation}
Its parameter $\mu^2$ determines the probability for a
cross-link to be formed, and therefore controls the sol-gel transition.
The average cross-link density is given by $ [M]/N = \mu^2/2$
and the standard deviation, relative to the mean, vanishes
in the thermodynamic limit. Thus, $\mu^2 = 2[M]/N$ is the
\emph{average coordination number}, i.e., the average number of
particles to which a certain particle is connected.  For a detailed
discussion of the Deam-Edwards distribution, see
\cite{RandomXlinkReview,FiniteDGelation}.

\section{Order parameter and free energy}
With help of the replica technique we obtain the disorder-averaged free energy 
$F = -[\ln Z] = -\lim_{n\to
  0}(\mathcal{Z}_{n+1}-\mathcal{Z}_{1})/(n\mathcal{Z}_{1})$.  The
replica partition function $\mathcal{Z}_{n+1}=\int{\cal D}\Omega
e^{-{\cal S}_\Omega}$ is represented as a functional integral over the
order-parameter field, whose expectation value
\begin{equation}
\Omega(\hat{x}) = \frac{1}{N} 
\sum_{i=1}^N \left\langle \delta({\bf x}^{(0)} - {\bf R}^{(0)}_i) \cdots
\delta({\bf x}^{(n)} - {\bf R}^{(n)}_i) \right\rangle_{{\cal S}_\Omega}
\end{equation}
quantifies the probability of finding a particle at position ${\bf x}^{(0)}$
in replica 0, at ${\bf x}^{(1)}$ in replica 1, etc. Here, the average $\langle \, \cdot \, \rangle_{{\cal S}_\Omega}$ is evaluated with the statistical weight $\e^{-{\cal S}_\Omega}$. To simplify the
notation, we have introduced \hbox{$d(n+1)$}-dimensional hatted
vectors $\hat{x} = ({\bf x}^{(0)}, ..., {\bf x}^{(n)})$.
Previous work \cite{DistOfLLength} has shown that the saddle-point solution for the order
parameter has the following simple form:
A fraction of monomers $(1-Q)$ is delocalized and can be found
anywhere in the sample with equal probability. The remaining fraction
$Q$ is the ``infinite" cluster and each particle $i$ performs
Gaussian fluctuations -- in each replica -- about a mean position
${\bf y}_{i}$ with localization length $\xi_i$. The mean positions ${\bf y}_{i}$ are
randomly distributed and the localization lengths $\xi$ follow the
distribution $\mathcal{P}(\xi^2)$. The resulting order parameter is
explicitly given by 
\begin{equation}
\Omega(\hat{x})=\frac{1-Q}{V^{n+1}}+\frac{Q}{V}\int_V \d^d y
\int_0^\infty \frac{\d \xi^2 \, \mathcal{P}(\xi^2)}{(2\pi \xi^2)^{dn/2}} 
\exp{\left(-\sum_{\alpha=0}^n \frac{({\bf x}_{\alpha}-{\bf y})^2}{2\xi^2}\right)} ,
\end{equation}
 with self-consistent equations for $Q$ and $\mathcal{P}(\xi^2)$ given
 in \cite{DistOfLLength,FiniteDGelation}. 
Note that the order parameter does not depend on $\sum_{\alpha=0}^n {\bf x}_{\alpha}$.

The Landau-Wilson free energy 
\begin{equation}
\label{effFreeE}
 {\cal S}_{\Omega} =\frac{ \mu^2 }{2}
  \overline{\sum_{\hat{k}}} \Delta(\hat{k}) \left| \Omega(\hat{k})
  \right|^2 - \ln{ \left(\int \frac{{\rm d}^{d(n+1)}R}{V^{n+1}} \, \exp \left(
    \mu^2 \overline{\sum}_{\hat{k}} \Delta(\hat{k})
      \Omega(\hat{k}) \e^{-i \hat{k} \hat{R} } \right) \right)}
\end{equation}
is particularly simple, because we consider cross-linked particles
instead of cross-linked chains. This will allow us to perform
computations in the \emph{highly cross-linked limit}.
The function $\Delta(\hat{k}) = \exp(-\frac{1}{2} a^2 \hat{k}^2)$
reflects the harmonic potential for the cross-links (\ref{HXlink}), and $\hat{k} := ({\bf k}^{(0)}, ... ,{\bf k}^{(n)})$. The Symbol $\overline{\sum}_{\hat{k}}$ means summation over all $\hat{k}$ with at least two non-zero components ${\bf k}^{(\alpha)}$.

\section{Spontaneously broken translational invariance}
\begin{figure}
\twoimages{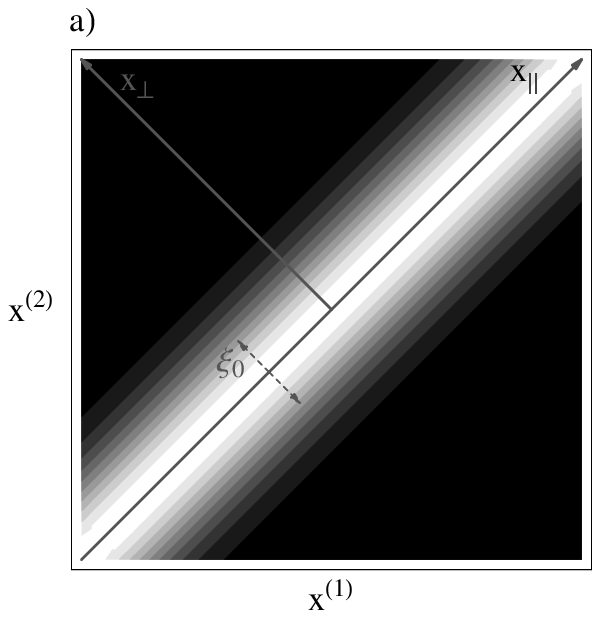}{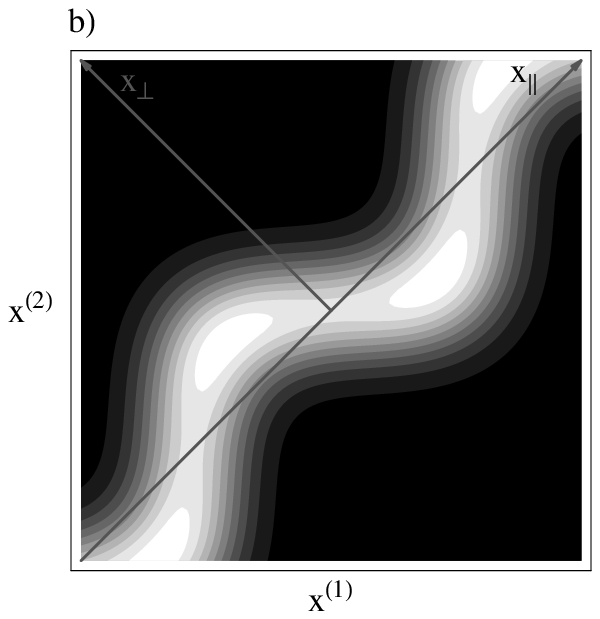}
\caption{Ansatz for the order parameter $\Omega(x^{(1)},x^{(2)})$ in
  real space for two replicas in one dimension. White indicates high
  values of $\Omega$. To simplify, we have assumed a fixed
  localization length $\xi_0$ in this figure. $\Omega(x^{(1)},x^{(2)})$ is the
  probability of finding a particle at position $x^{(1)}$ in replica $1$
  and at $x^{(2)}$ in replica $2$. For the rotated coordinate system
  $x_\parallel = (x^{(2)} + x^{(1)})/\sqrt{2}$ and $x_\bot = (x^{(2)}
  - x^{(1)})/2$. In (a) the displacement vector $u_\bot \equiv 0$, and
  in (b) $u_\bot = \sin x_\parallel$. (Note, however, that for physical
  systems typically $\xi \ll$ wavelength of $u_\bot$.) }
\label{fig:OrderParameter}
\end{figure}

The effective replicated Hamiltonian is invariant under uniform
translations of all particles, separately in each replica, i.e., the
transformation ${\bf R}_i^{\alpha}\rightarrow {\bf R}_i^{\alpha}+{\bf
u}^{\alpha}$ is an exact symmetry of the Hamiltonian\footnote{~The effective replicated Hamiltonian is also invariant under uniform rotations, separately in each replica. However the spontaneous breaking of this symmetry does not give rise to Goldstone modes and is not discussed further here.}. In the amorphous solid phase, which is characterized by random
localisation of a finite fraction of particles, this symmetry is
spontaneously broken, and the only symmetries remaining are common
displacements of all replicas, i.e., ${\bf R}_i^{\alpha}\rightarrow
{\bf R}_i^{\alpha}+{\bf u}$, reflecting the macroscopic translational
invariance of the amorphous solid phase. (For a detailed discussion see\cite{GoldstoneFluct,GoldstoneFluctEPL}.) Consequently there is a whole
family of order parameters that each give rise to the same free
energy and are related by replica-dependent, spatially uniform
translations. To distinguish common and relative displacements of the
replicas, it is helpful to split up the \hbox{$d(n+1)$}-dimensional
hatted vectors $\hat{x}$ into $d$-dimensional longitudinal ones ${\bf
  x}_\parallel = \frac{1}{\sqrt{n+1}} \sum_{\alpha=0}^n {\bf
  x}^{(\alpha)}$ and $dn$-dimensional transverse ones $x_\bot =
\hat{x} - \frac{1}{\sqrt{n+1}} ({\bf x}_\parallel,...,{\bf
  x}_\parallel)$. The manifold of symmetry-related saddle points is
parameterized by $u_\bot = \hat{u} - \frac{1}{\sqrt{n+1}} ({\bf
  u}_\parallel,...,{\bf u}_\parallel)$:
\begin{equation}
 \Omega_{u_{\bot}}(\hat{x})=\frac{1-Q}{V^{n+1}}+\frac{Q}{V}
\int_0^\infty \frac{\d \xi^2 \, \mathcal{P}(\xi^2)}{(2\pi \xi^2)^{dn/2}} 
\exp \left(-\frac{(x_\bot - u_\bot)^2}{2 \xi^2} \right).
\end{equation}
The situation is illustrated in fig.~\ref{fig:OrderParameter}a, where
the order parameter is shown for two replicas in one space-dimension.
The order parameter does not depend on ${\bf
  x}_\parallel$, so that it is represented by a rectilinear ridge having its
maximum height along the axis $x_\bot = 0$ and having width $\xi$.
The family of symmetry-related solutions is generated by rigidly translating
the ridge to a new position: $x_\bot = u_\bot$.
(The delocalized particles contribute only an unimportant
constant background $(1-Q)/V^{n+1}$, which has been ignored in
the figure.)

\section{Goldstone fluctuations}
Low energy excitations are constructed by particle displacements
$u_\bot({\bf x}_\parallel)$ that depend on the position in the
sample\cite{GoldstoneFluctEPL,GoldstoneFluct}. The ridge is not
translated rigidly, but deformed by the position-dependent vector
$u_\bot({\bf x}_\parallel)$, as indicated in
fig.~\ref{fig:OrderParameter}b. These excitations \emph{do} cost
energy; only in the long wavelength limit do we expect the energy of
excitation to vanish with an inverse power of the wavelength.  The
generalized order parameter fluctuations take the form \cite{mistake}:
\begin{equation}
\label{ansatzX}
\Omega_{u_\bot} (\hat x) = 
\frac{1-Q}{V^{n+1}} +  \frac{Q}{V(n+1)^{d/2}} 
\int_0^\infty \frac{\d \xi^2 \, \mathcal{P}(\xi^2)}{(2\pi \xi^2)^{d(n+1)/2}} 
\; \int \d^d y_\parallel e^{-\frac{({\bf y}_{\shortparallel} - {\bf x}_\shortparallel )^2 + 
(x_\bot - u_\perp({\bf y}_{\shortparallel}))^2}{2 \xi^2} } \; .
\end{equation}
We only consider pure shear deformations, and hence require that $|\det (\delta_{\nu,\mu} + \partial_\nu u_\mu^{\alpha})|=1$. This constraint
guarantees the preservation of volume, and simplifies for small deformations to
$\partial_\nu u_\nu^{\alpha}=0$.

We insert the Ansatz (\ref{ansatzX}) into the general expression
(\ref{effFreeE}) for the free energy and only consider small
distortions, so that it is sufficient to keep $u_\bot$ only to the lowest
order, $( \partial u_\bot / \partial {\bf x}_\parallel )^2$. Higher
orders in $u_\bot$ and higher-order derivatives are neglected. Restoring units of energy, the resulting free energy is 
\alpheqnc
\begin{equation}
\label{elastic}
{\cal S}_{u_\bot}={\cal S}_{u_\bot=0}+ 
\frac{G}{2} %\left(
\lim_{n \to 0} \frac{1}{n} \int {\rm d}^d {x}_\parallel 
\sum_{\alpha} \sum_{\nu, \mu}
\left( \frac{\partial u_{\bot \nu}^{(\alpha)}}{\partial x_{\parallel \mu}} \right)^{\!\!2} %\frac{\partial u_{\bot \nu}^{(\alpha)}}{\partial x_{\parallel \mu}} \right)
,
\end{equation}
with the shear modulus $G$ given by
\begin{equation}
  G = \left( \mu^2 Q - 1 + \exp(-\mu^2 Q) - 
\frac{\mu^2  Q^2}{2} \right) n_0 \, k_{\rm B} T \, ,
    \label{shearMod}
\end{equation}
\reseteqnc
where $n_0 = N/V$ is the mean particle-density.

\begin{figure}
 \oneimage{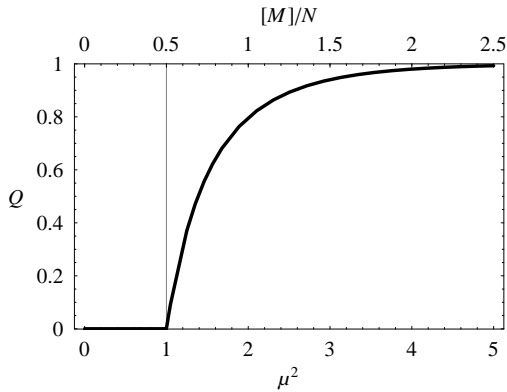}
 \caption{Fraction of localized particles $Q$, depending on the average
   coordination number. The infinite cluster forms at the critical value
   $\mu_c^2 = 1$.}
 \label{fig:Q(mu)}
\end{figure}

The part of the Landau-Wilson free energy (\ref{elastic}) that does not depend on $u_\bot$ determines the gel fraction and the
distribution of localization lengths that was discussed in
\cite{FiniteDGelation}. In particular, differentiating $ {\cal S}_{u_\bot=0}$
with respect to $Q$ yields a simple relation between the gel fraction
$Q$ and the average coordination number $\mu^2$:
$\exp(-\mu^2 Q) = 1 - Q$, which is shown in fig.~\ref{fig:Q(mu)}. 
It was previously derived in many other contexts
\cite{FiniteDGelation,DistOfLLength,RandomXlinkReview}. 

\section{Shear modulus}
In eq.~(\ref{shearMod}), we identified the shear modulus $G$ in the
free energy. As one would expect, it vanishes in the liquid phase
($\mu^2 < 1$). In the solid phase ($\mu^2 > 1$), it
increases linearly with particle density $n_0$ and depends only on the
cross-link density, given the gel fraction $Q=Q(\mu^2)$.

\begin{figure}
\twoimages[viewport= 20 0 220 150,clip]{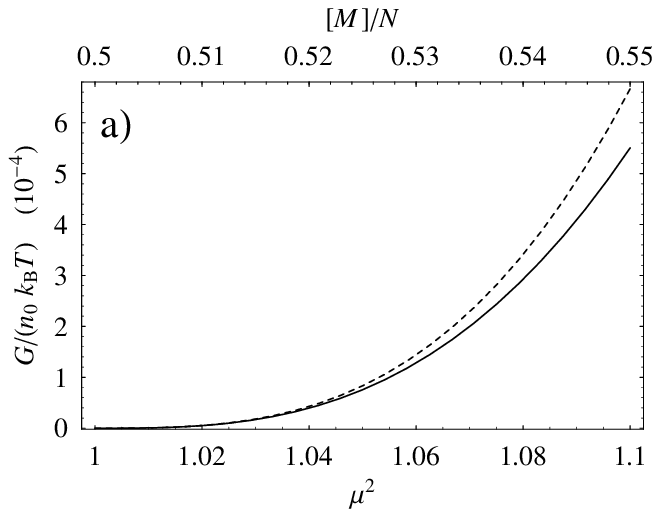}{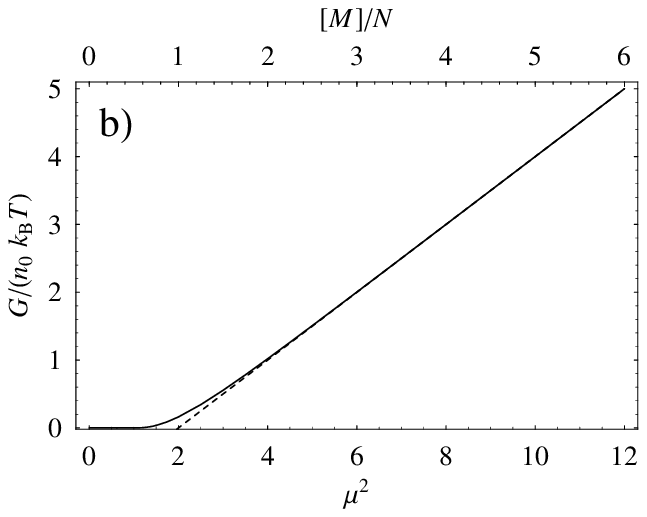}
\caption{Shear modulus $G$ divided by the particle
  density $n_0$ and $k_{\rm B}T$, depending on the average coordination number
  $\mu^2$. The solid line shows the exact expression
  (\ref{shearMod}), the dashed lines are the approximations valid (a) close
  to the sol-gel transition, eq.~(\ref{shearModSmall}), and (b) for
  highly connected systems, eq.~(\ref{shearModBig}).}
\label{fig:shearMod}
\end{figure}

Close to the gel transition \hbox{$\mu^2 \gtrsim 1$}, the
shear modulus can be expanded in terms of \hbox{$\epsilon := \mu^2
  -1$}, % = 2([M]/N - 1/2) $,
resulting in power-law behavior (see fig.~\ref{fig:shearMod}a):
\begin{equation}
G = \frac{2}{3} n_0 \, k_{\rm B} T \, \epsilon^3 + O(\epsilon^4) \; .
\label{shearModSmall}
\end{equation}
For the highly-connected regime, $\mu^2 \gg 1$, %$Q \approx 1$ and 
$G$ reduces to (see fig.~\ref{fig:shearMod}b):
\begin{equation}
G = \frac{1}{2} n_0 \, k_{\rm B} T \, (\mu^2 - 2) + O(\mu^{-2}) \; .
\label{shearModBig}
\end{equation}

The scaling $G \sim \epsilon^3$ close to the critical point has been derived previously \cite{GoldstoneFluctEPL,GoldstoneFluct}; the critical exponent is a result of mean field theory and expected to be modified by interacting fluctuations below $d=6$ \cite{PengRenormalization}. In the highly cross-linked limit, fluctuations are presumably weak so that our result is expected to hold even beyond the Gaussian expansion around mean field theory. The result we get is reminiscent of the classical theory of rubber elasticity, $G=n_{\rm s} k_{\rm B} T$, with the density of strands $n_{\rm s}$ being replaced by the density of cross-links: $\frac{1}{2} \mu^2 n_0 = [M]/V$.

In order to compare the calculated shear modulus with experimental results, note that, if the cross-link length is small compared to the particle size, it is unlikely for one monomer in $d=3$ ($d=2$) dimensions to be connected to more than 12 (6) other monomers. Therefore, average coordination numbers $\mu^2 > 12$ ($>6$) are inaccessible in these systems. Only if long tethers can form, reaching further than the nearest neighbour, higher average coordination numbers $\mu^2$ are possible.

\section{Conclusions}
We have determined the shear modulus $G$ of a randomly
cross-linked system. It only depends on the particle density $n_0 =
N/V$ and the average coordination number $\mu^2$.
Interestingly, neither the distribution of localization lengths
$\mathcal{P}(\xi^2)$, which entered the order parameter in
eq.~(\ref{ansatzX}), nor the typical cross-link length $a$ influence
the macroscopic behavior of the material. This result is consistent
with the observation that the shear modulus determines the response of
the system to a shear deformation on the longest scales: The
wavelength of the shear deformation has to be larger than all
microscopic length scales. Furthermore, one observes that $G \sim T$. Therefore, the shear elasticity is purely
entropic. In contrast, the bulk modulus should depend
on the strength of the excluded-volume interaction, and hence it is not
expected to be as universal as the shear modulus.

We also stress that the shear modulus does not depend on the spatial
dimension $d$, apart from the fact that the maximal coordination
numbers depend on dimensionality and determine the physical range of
$\mu^2$. Again, this could be expected, because we only consider a
Gaussian expansion around the saddle point; interactions between fluctuations will most likely cause a dependence on dimensionality.

It is also of interest to study the local elasticity, i.e., the response
to shear deformations of finite or even small wavelength. Random
networks are strongly inhomogeneous, so that one expects the
elasticity to fluctuate spatially. Work along these lines is in progress.

\acknowledgments
This work was supported by the DFG through SFB 602 and Grant No. Zi 209/7
(S.U,A.Z) and by the U.S.~National Science Foundation through grants DMR 02-0858 and 06-08516 (X.M.,P.M.G.).


\begin{thebibliography}{0}

\bibitem{DeamEdwards}
  \Name{Deam R.~T.~\and Edwards S.~F.}
  \REVIEW{Phil. Trans. R. Soc. A}{280}{1976}{317}.

\bibitem{Ball}
  \Name{Ball R.~C. \and Edwards S.~F.}
  \REVIEW{Macromolecules}{13}{1980}{748}.

\bibitem{RandomXlinkReview}
  \Name{Goldbart P.~M., Castillo H.~E.~\and Zippelius A.}
  \REVIEW{Adv. Phys.}{45}{1996}{393}.

\bibitem{Panyukov}
  \Name{Panyukov S.~V. \and Rabin Y.}
  \REVIEW{Phys. Rep.}{269}{1996}{1}.

\bibitem{GoldstoneFluctEPL}
  \Name{Mukhopadhyay S., Goldbart P.~M.~\and Zippelius A.}
  \REVIEW{Europhys. Lett.}{67}{2004}{49}.

\bibitem{GoldstoneFluct}
  \Name{Goldbart P.~M., Mukhopadhyay S.~\and Zippelius A.}
  \REVIEW{Phys. Rev. B}{70}{2004}{184201}.

\bibitem{FiniteDGelation}
  \Name{Broderix K., Weigt M.~\and Zippelius A.}
  \REVIEW{Eur. Phys. J. B}{29}{2002}{441}.

\bibitem{Zallen}
  \Name{Zallen R.}
  \Book{The Physics of Amorphous Solids}
  \Publ{John Wiley}
  \Year{1983}.

\bibitem{DistOfLLength}
  \Name{Castillo H.~E., Goldbart P.~M.~\and Zippelius A.}
  \REVIEW{Europhys. Lett.}{28}{1994}{519}.

\bibitem{mistake}
  In Refs.\cite{GoldstoneFluctEPL,GoldstoneFluct}
   a slightly different ansatz for $\Omega(\hat{x})$ was used, which
   was not quite correct, but gave rise to the same scaling $G \sim
   \epsilon^3$.

%\bibitem{Universality}
%  \Name{Peng W., Castillo H.~E., Goldbart P.~M.~\and Zippelius A.}
%  \REVIEW{Phys. Rev. B}{57}{1998}{839}.

\bibitem{PengRenormalization}
  \Name{Peng W. \and Goldbart P.~M.}
  \REVIEW{Phys. Rev. E}{61}{2000}{3339}.

\end{thebibliography}
\end{document}